\pdfoutput=1

\documentclass[11pt]{article}

\usepackage[final]{acl}

\usepackage{times}
\usepackage{latexsym}
\usepackage{kotex}
\usepackage{graphicx} 
\usepackage{url}

\usepackage[T1]{fontenc}

\usepackage[utf8]{inputenc}

\usepackage{microtype}

\usepackage{inconsolata}

\title{Three Disclaimers for Safe Disclosure: \\A Cardwriter for Reporting the Use of Generative AI in Writing Process}

\author{Won Ik Cho\thanks{Equal Contribution.} \\
  Seoul National University\thanks{Done after graduation.} \\
  \texttt{tsatsuki@snu.ac.kr} \\\And
  Eunjung Cho\textsuperscript{\textasteriskcentered} \\
  ETH Zürich \\
  \texttt{choeun@ethz.ch} \\\And
  Hyeonji Shin\textsuperscript{\textasteriskcentered} \\
  Seoul National University \\
  \texttt{wl7788@snu.ac.kr } \\}

\begin{document}
\maketitle
\begin{abstract}
Generative artificial intelligence (AI) and large language models (LLMs) are increasingly being used in the academic writing process. This is despite the current lack of unified framework for reporting the use of machine assistance. In this work, we propose "Cardwriter", an intuitive interface that produces a short report for authors to declare their use of generative AI in their writing process. The demo is available online\footnote{\url{https://cardwriter.vercel.app}}.
\end{abstract}

\section{Introduction}

With recent advancements in generative artificial intelligence (AI) such as text-to-image, audio systems, and large language models, we are facing a new type of intellectual property that leverages these technological advances \cite{hbr2023generative}. This includes not only artworks such as paintings and music, but also various forms of writings and codes (programming languages with semantics, syntax, and pragmatics), and more.

However, there exists a gap between technological advances and the maturity of the level of public acceptance. As a result, we often see public backlash when people find out that machine assistance was used in creating a piece of work \cite{nyt2022generative}, or when the use was not explicitly declared \cite{wong2024ai}.
For artworks, it has become a usual practice to declare if/which machine assistance was used to create them \cite{adobe2023generative}. This is also because model-generated artworks like paintings or music are often not modified after generation if the user does not have the relevant expertise in the art/music domain. However, for writings and codes - a sequence of discrete symbols -, in most cases, some edits are made before publishing the work. The author(s) of the work usually take the authorship without explicitly declaring the use of machine assistance. While this convention is largely due to the prohibition of listing the machine as an author \cite{thorp2023chatgpt}, if relevant disclaimers are not appropriately made, some unfortunate mishaps could harm the authors' reputation and academic integrity (See \citet{zhang2024three} and the beginning line of Introduction). 

Despite such possibilities, simply not using any machine assistance in the  writing process is not practical. Especially in modern days where English has become a de facto language in academic writing, using assistance of not only generative AI but also tools like Grammarly\footnote{\url{https://www.grammarly.com/}}, Quillbot\footnote{\url{https://quillbot.com/}}, Google Translate\footnote{\url{https://translate.google.com/}} or DeepL\footnote{\url{https://www.deepl.com/translator/}} are inevitable for improving the quality of writing \cite{khabib2022introducing}. However, adopting generative AI is different compared to using simple editing tools, since generative AI can generate information that does not exist in the user input. This makes it challenging to distinguish the intellectual contributions made by humans from that by the machine, and there is no clear-cut answer yet to this concern. 
Although academic writing is an area where transparency and integrity are important, there are no clear guidelines yet on how and to what extent authors should disclose their use of generative AI.

Against this backdrop, we propose a tool that enables authors to transparently share whether they have used machine assistance - especially generative AI - in their writing process. The theoretical backing of this demonstration is outlined in our previous study \cite{cho2023papercard}, where we discuss the need, and suggest a systematic method, for a transparent disclosure of machine assistance used. 
Our main contributions through this demonstration are as follows:
\begin{itemize}
    \item We propose Cardwriter, a system that streamlines and automates generation of PaperCard -- a documentation for transparent reporting of machine use in academic writing. 
    \item By proposing the system, we contribute to building a healthy culture of using machine assistance in academic writing.
\end{itemize}

\begin{figure}
    \centering
    \includegraphics[width=\columnwidth]{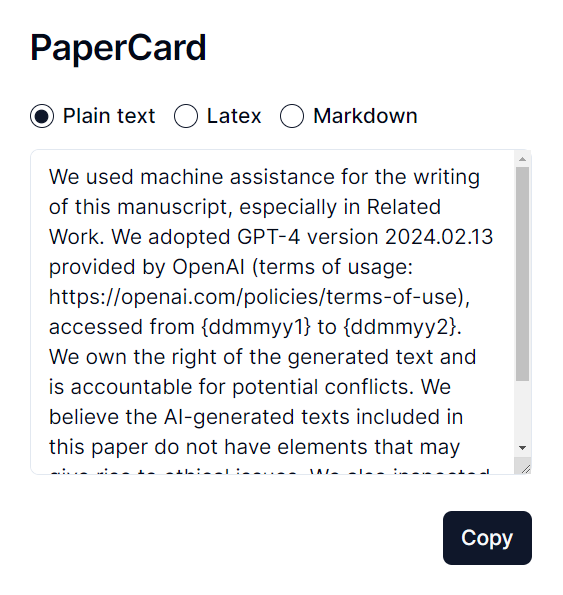}
    \caption{A Sample PaperCard produced with Cardwriter}
    \label{fig:card}
\end{figure}

\section{Background}


In academia, authorship is generally defined as the recognition of an individual’s intellectual contribution to a piece of academic work \cite{osborne2019authorship}. Some of the most common criteria for meriting authorship include: 1) the author(s) have made significant intellectual contributions to the work; 2) have been involved in the planning, execution, analysis, interpretation, writing and reviewing of the manuscript, and 3) have approved the final version of the manuscript. The importance of authors being able to take public responsibility for the integrity of the work is also often emphasised.

While technological developments in the past, such as typewriter, computational memory and graphical user interfaces, the Internet, and AI-supported text editors or translators have also had a great impact on the process of academic writing, the challenge they posed to the idea of authorship was relatively limited. However, with the advent of generative AI that can generate texts that are indistinguishable from human-generated texts, the challenge it poses to the idea of authorship becomes more nuanced. Researchers found various ways to use generative AI to assist them with writing their academic papers, from selecting topics for the manuscript \cite{srivastava2023day}, writing specific sections of the manuscript \cite{aydin2022openai}, to producing an entire manuscript \cite{frye2022should}. Some even merited authorship to the generative AI used in their writing process \cite{frye2022should}. However, the use of these models in academic writing raises various concerns, including the potential for a decrease in the originality and creativity of academic papers \cite{baron2021know,baron2023chatgpt}, the accuracy and reliability of the text generated by these models. Despite these concerns, the use of these models in academic writing is likely to grow as researchers continue to explore their potential applications.

There have recently been active discussions within academia on how to govern the use of such tools. Some argue for allowing the use of AI-generated content while imposing restrictions on how it is used. For example, Nature announced that researchers using large language models such as ChatGPT should clearly document the use in appropriate sections such as methods or acknowledgements sections \cite{editorials2023tools}. Others outright ban the use of AI-generated content, as seen in the recent guidelines published by International Conference on Machine Learning \cite{icml2023clarification} and \citet{science2023clarification}. A number of universities are considering revisions to their academic integrity policies so their plagiarism definitions include generative AI. There are also various ongoing efforts to develop technological solutions to detect AI-generated texts \cite{mitchell2023detectgpt,wiggers2022openai}.

This highlights the need for a systematic way to transparently report the use of such tools. However, requiring authors to submit a separate report can not only be a burden to the authors, but also to the reviewers. A system that streamlines and automates generation of PaperCard -- as in Figure~\ref{fig:card} -- would be useful, similar to the Model Card Toolkit \cite{tf2021mct} for generating Model Cards \cite{mitchell2019model}.


\begin{figure}
    \centering
    \includegraphics[width=0.8\columnwidth]{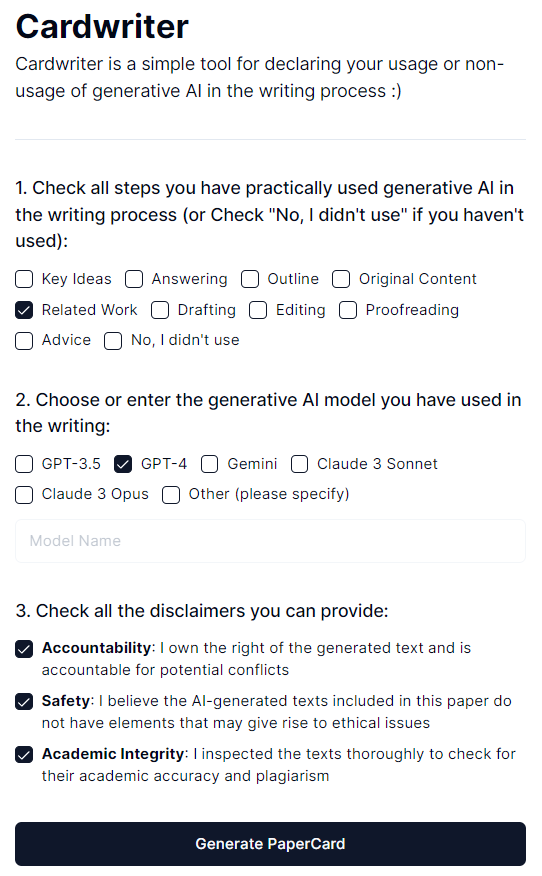}
    \caption{User interface (UI) of Cardwriter}
    \label{fig:user_interface}
\end{figure}

\begin{figure*}
    \centering
    \includegraphics[width=\textwidth]{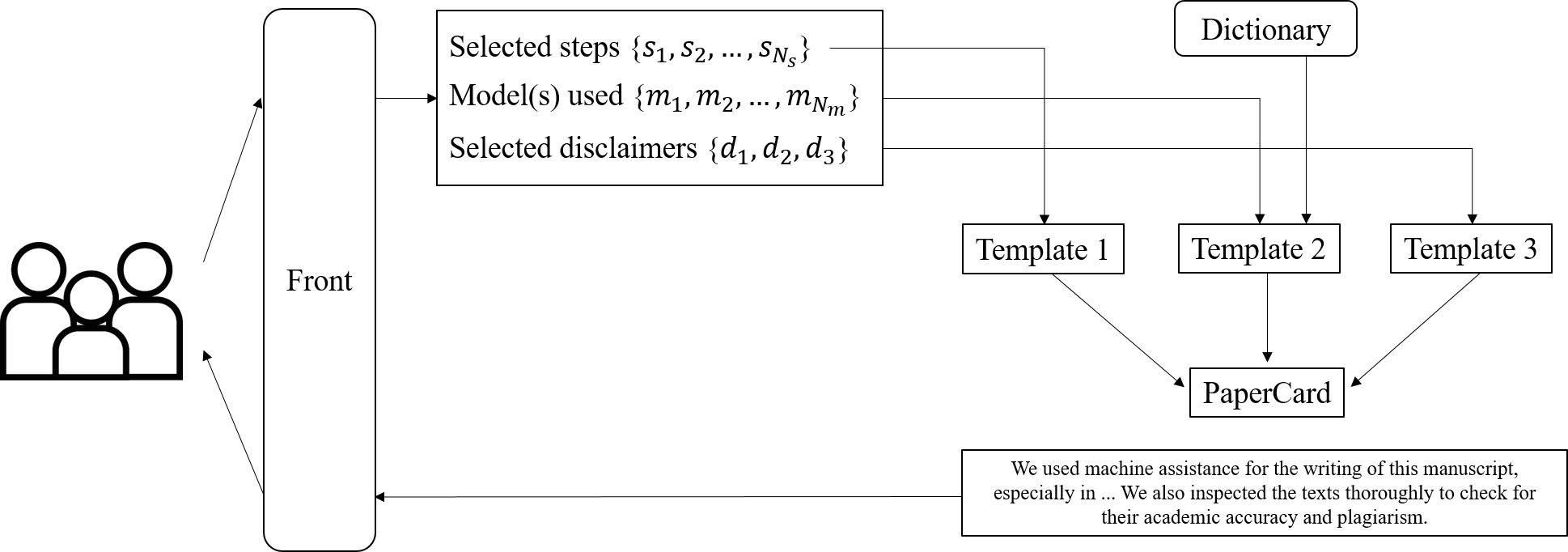}
    \caption{Data-flow diagram}
    \label{fig:processor}
\end{figure*}

\section{System Overview}

Our proposed system consists of the following three components: 

\begin{enumerate}
    \item A front-end that receives user requests (see Figure~\ref{fig:user_interface}) 
    \item A processor that receives user input and generates the body of a PaperCard (See Figure~\ref{fig:processor})
    \item A front that displays the PaperCard in a format ready for users to copy into their work (see Figure~\ref{fig:card})
\end{enumerate}

While the system can be used for other domains such as code writing, its primary target is academic writing domain.

\subsection{User Interface}
Users first select how they used generative AI in their writing process (Part 1 of Figure~\ref{fig:user_interface}). Descriptions for each checkbox pops up as a toggle when the user places their cursor on the checkbox. The descriptions are based on those provided in the original PaperCard paper \cite{cho2023papercard}. 
For users who did not use generative AI in their writing process can also declare so, by checking the last checkbox. The output will then simply be \textit{"The authors did not use any assistance from generative AI in writing this manuscript."}

The user then chooses the generative AI model(s) used (GPT-3.5\footnote{\url{https://openai.com/chatgpt}}, GPT-4 \cite{achiam2023gpt}, Gemini \cite{team2023gemini}, Claude 3 Sonnet, and Claude 3 Opus\footnote{\url{https://claude.ai/chat/}}) or by direct typing (Part 2 of Figure~\ref{fig:user_interface}). Only commercial LLMs serviced by big tech companies are currently listed, but models that are not in the list can also be entered. For such custom cases, the model with the most similar name in the dictionary will be provided, and the user can edit the information later. 

Finally, the user selects three disclaimer checkboxes (Part 3 of Figure~\ref{fig:user_interface}) that help encourage them to understand and consider issues regarding the license of the generated information by the machine, whether using the information is safe, and whether they would guarantee the user's academic integrity.

\subsection{Processor}
In the processor, the expressions to be entered in the card are completed using user input conveyed from the UI. Since the message does not need to be lexically fluent, we adopted a template to complete the expressions\footnote{If a different expression is preferred, generative AI could be utilised in future system modification.}. The three types of information received from the UI are processed separately, with steps 1 and 3 completed via a simple template fill-in, and step 2 through dictionary matching (Figure~\ref{fig:processor}). 

Let the following be a user input:

\begin{itemize}
    \item $S = \{s_1, s_2, ... , s_{N_s}\}$: selected steps
    \item $M = \{m_1, m_2, ... , m_{N_m}\}$: selected models 
    \item $D = \{d_1, d_2, d_3\}$: indicators of disclaimers that need to be checked (all Boolean)
\end{itemize}

The scheme for matching user's manual input with dictionary is executed via exact matching, but would be further improved with similarity checks.

In steps 1 and 3, the following template is adopted:

\begin{itemize}
    \item Step 1: “\textit{We used machine assistance for the writing of this manuscript, especially in <all steps selected from $S$>.}”
    \item Step 3: “\textit{<if $d_1$ is True> We own the rights of the generated text and are accountable for potential conflicts. <if $d_2$ is True> We believe the AI-generated texts included in this paper do not have elements that may give rise to ethical issues. <if $d_3$ is True> We inspected the texts thoroughly to check for their academic accuracy and plagiarism.}”
\end{itemize}

\paragraph{Dictionary: source of information}
Dictionary currently contains manually collected information stored separately as \textit{json}, and includes the model's official name, the provider, service URL, the URL of terms of use, and the most recent version (date). The dictionary is managed in a form that can be updated later manually or via community contributions. The following is a sample \textit{json} of GPT-4 model.\medskip

\textit{\{"model": "GPT-4", "provider": "OpenAI", "url": "https://chat.openai.com/", "terms": "https://openai.com/policies/terms-of-use", "version": "2024.02.13"\}}\medskip

Given the dictionary, after the matching of user input $m$ and the ``\textit{model}" attribute, the processor yields the body of step 2 using the following template:

\begin{itemize}
    \item Step 2: \textit{“We adopted <model> (url: \url{<url>}) version <version> provided by <provider> (terms of usage: \url{<terms>}), accessed from \{ddmmyy1\} to \{ddmmyy2\}.”}
\end{itemize}

where \textit{\{ddmmyy1\}} and \textit{\{ddmmyy2\}} can be customised by the user to exhibit the date of access.


\subsection{Display}
The texts produced by the processor are aggregated, and the final body is sent back to the front-end. The output text corresponding to steps 1-3 are displayed in a single text box (Figure~\ref{fig:card}). The user can copy the content of the card by selecting a format among the choices of plain text, LaTeX, and Markdown. 

\section{Demonstration}

Based on the system properties described above, we explain the intended use of the system and demonstrate the user guidance.

\subsection{Intended Use}

Users who use PaperCards through Cardwriter can generally have the following purposes.

\begin{itemize}
    \item All authors who did not use generative AI in the academic writing process (but who want to declare that they did not use the assistance)
    \item All authors who used a commercial or open-source LLM in academic writing (and that they want to declare the machine assistance)
    \item All authors of non-academic writing who did or did not use the machine assistance in producing their outcome (including blog articles, essays, school assignments, etc.)
\end{itemize}

For these authors, using our platform can be an intuitive and accessible choice that one can easily get the structured declaration format to be provided to the publishers or to audience.

\subsection{User Guidance}

To achieve the above goals, users can access the system to modify and distribute the results. The output can be inserted into the manuscript's acknowledgment, ethical statement, or can be described as a separate section. If a user needs to follow a specific guideline for declaring their use of machine assistance in the writing process, the user can take the system output as a foundation and tailor it to fit the guideline. There is no specific terms of use for our proposed system, though users can declare that the card was produced by our web demo or following the original paper that proposes the concept \cite{cho2023papercard}.

For a better user experience and further maintenance, we provide the code and guideline via Github repository\footnote{\url{https://github.com/nyanxyz/cardwriter}}. The repository contains a brief summary of the project, how to use, and a link to short video\footnote{Link available in the Github README.} demonstrating the sample usage of the system.

\section{Discussion}
Our motivation for this system demonstration primarily stems from recent concerns about using machine assistance - especially generative AI - in academic writing, where transparent reporting of the use would be be helpful for authors' academic integrity. Despite the intuitive implementation of the proposed system, our approach has a few limitations.

\subsection{Limitations}

One major limitation of a user-oriented declaration scheme is that it is completely up to the integrity of the authors. That is, unless a comprehensive inspection can be conducted for the authors' commercial generative AI server logs, readers, reviewers, and editors are unable to determine whether the author(s) have used any form of machine assistance in their writing process. This is why sometimes a seemingly AI-generated response is inadvertently included in the writing through copy-paste, yet it is difficult to be conclusive that the authors have done so from simply reading the paper (since making assumptions also yields undesirable side effects). 

In addition, our approach requires a continuous management of the system, which may require human resources to regularly update the model candidates and implement better system for matching model names. This is why we welcome community recognition and contribution of our project. In a similar vein, as the scope of machine assistance in the academic writing process continues to grow -- i.e. help with creating figures, tables, and supplementary materials --, the PaperCard's coverage should also grow. We leave this for future work.

\subsection{Societal Implications}

Despite the limitations, our system has positive societal implications that can be directly acknowledged by the academic community. First, it provides a practical solution to the current situation where there is no clear framework for authors to make disclaimers. Although it is up to the authors' academic integrity for making the disclaimers, the existence of an accessible framework the authors can easily adopt can reduce the burden. Also, while the generation system is currently provided only as a web demo, the code will be open-sourced at later stage to facilitate community contribution. This can accelerate the development and expansion of the system, not only to other aspects of academic writing process, but also to other domains such as code writing.


\section{Conclusion}

In this work, we introduced a system demonstration for authors to conveniently report their use of machine assistance (specifically generative AI) in their academic writing process. The motivation behind this work has been explained in detail in Introduction and Background sections, and the system consisting of UI, processor, and display is shown with explanations for intended use, as well as a brief user guideline. We hope our system can be a cornerstone for establishing a culture of transparent declaration of the use of generative AI. We also hope our system can be further improved with public recognition and community contribution.

\section*{Ethical Statement}

We declare that no financial aid was received in conducting this study and that the research was done without any relationship that could be construed as a potential conflict of interest. No study on human participants, or experiment that requires disclosure of data, processes, or results was conducted. Also, the authors did not use any assistance from generative AI in writing this manuscript.

\section*{Acknowledgements}

Authors thank Kyunghyun Cho for advising the original direction of this project. Authors also appreciate all the community and researchers who provided constructive feedback in driving the implementation.

\bibliography{custom}





\end{document}